\documentclass[a4paper, amsfonts, amssymb, amsmath, reprint, showkeys, nofootinbib, twoside]{revtex4-2}
\usepackage[english]{babel}
\usepackage[utf8]{inputenc}
\usepackage[colorinlistoftodos, color=green!40, prependcaption]{todonotes}
\usepackage{amsthm}
\usepackage{mathtools}
\usepackage{physics}
\usepackage{xcolor}
\usepackage{graphicx}
\usepackage[left=13mm,right=13mm,top=35mm,columnsep=15pt]{geometry} 
\usepackage{adjustbox}
\usepackage{placeins}
\usepackage[T1]{fontenc}
\usepackage{lipsum}
\usepackage{csquotes}
\AtBeginDocument{\RenewCommandCopy\qty\SI}
\usepackage[pdftex, pdftitle={Article}, pdfauthor={Author}]{hyperref} 
\usepackage{braket}
\usepackage{siunitx}

\usepackage{xcolor}  
\usepackage{textcomp}
\usepackage{gensymb}
\DeclareSIUnit\bar{bar}
\DeclareSIUnit\cps{cps}
\DeclareMathOperator{\sinc}{sinc}

\usepackage{nameref}

\begin{document}

\author{Anica Hamer}
    \affiliation{Physikalisches Institut, Rheinische Friedrich-Wilhelms-Universität Bonn, Bonn, Germany}

\author{Frank Vewinger}
    \affiliation{Institut für Angewandte Physik, Rheinische Friedrich-Wilhelms-Universität Bonn, Bonn, Germany}

\author{Thorsten Peters}
    \affiliation{Technische Universität Darmstadt, Darmstadt, Germany}

\author{Michael H. Frosz}
    \affiliation{Max-Planck-Institut für die Physik des Lichts, Erlangen, Germany}

\author{Simon Stellmer}
    \email[Correspondence email address: ]{stellmer@uni-bonn.de}
    \affiliation{Physikalisches Institut, Rheinische Friedrich-Wilhelms-Universität Bonn, Bonn, Germany}

\date{December 3, 2024} 

\title{Frequency conversion in a hydrogen-filled hollow-core fiber using continuous-wave fields}

\begin{abstract}
In large-area quantum networks based on optical fibers, photons are the fundamental carriers of information as so-called flying qubits. They may also serve as the interconnect between different components of a hybrid architecture, which might comprise atomic and solid state platforms operating at visible or near-infrared wavelengths, as well as optical links in the telecom band. Quantum frequency conversion is the pathway to change the color of a single photon while preserving its quantum state. Currently, nonlinear crystals are utilized for this process. However, their performance is limited by their acceptance bandwidth, tunability, polarization sensitivity, as well as undesired background emission. A promising alternative is based on stimulated Raman scattering in gases.

Here, we demonstrate polarization-preserving frequency conversion in a hydrogen-filled anti-resonant hollow-core fiber. This approach holds promises for seamless integration into optical fiber networks and interfaces to single emitters. Disparate from related experiments that employ a pulsed pump field, we here take advantage of two coherent continuous-wave pump fields.
\end{abstract}



\maketitle

\section{Introduction}

Highly efficient and state-preserving single-photon frequency conversion is a prerequisite for the realization and operation of quantum computing and communication in a hybrid architecture with optical interconnections \cite{Kimble2008,Silberhorn2017,Becher2012,Krutyanskiy2017,Fisher2021,Bell2017,Zhou2014,Ikuta2011,DeGreve2012,Tyumenev2022-sp,Winzer:18,Deutsch2023}. The two most well-known methods for quantum frequency conversion rely on the $\chi^{(2)}$ nonlinearity of crystals \cite{Silberhorn2017, Becher2012, Fisher2021, Bell2017, Zhou2014, Ikuta2011,DeGreve2012} and the $\chi^{(3)}$ nonlinearity of atomic gases in combination with short-pulse lasers \cite{Eramo1994,Tamaki1998,Babushkin2008,Cassataro:17}.

Developments based on the $\chi^{(2)}$ nonlinearity of crystals have reached near-unit conversion efficiency and a drastically reduced incoherent background. Still, the $\chi^{(3)}$ nonlinearity of molecular gases offers many benefits, as it sidesteps some of the limitations inherent to crystals, among them a very narrow acceptance bandwidth, polarization-dependence of the conversion process, absorption, heat management, and background emission.

In recent work, coherent Stokes and anti-Stokes Raman scattering (CSRS/CARS) processes based on the $\chi^{(3)}$ nonlinearity in gases have been demonstrated for frequency conversion \cite{10.1063/1.5030335, Shipp2017RamanST, Tian2023, BOYD20081}. These processes are intrinsically broadband, insensitive to the input polarization, and can operate on a wide selection of molecules and their transitions \cite{Tyumenev2022-sp,Aghababaei2023-pa, 10.1063/1.5111000}.  

In this Letter, we present frequency conversion from \SI{863}{\nano\meter} to \SI{1346}{\nano\meter} based on a CSRS process inside an antiresonant-reflecting hollow-core fiber (ARR-HCF). While the process itself has already been established \cite{Aghababaei2023-pa,Hamer:24}, we here show that the mode confinement increases the conversion efficiency. Remarkably, the fiber does not deteriorate the polarization, and interaction with the glass structure does not induce undesired Raman scattering. To our knowledge, all related experiments to date employed pulsed lasers to generate the coherence. By contrast, here we employ a pair of precisely adjusted continuous-wave (cw) lasers to drive the conversion process. This approach broadens the scope of applications to cases where the arrival time of the probe photon is unknown. Importantly, applications involving atoms or ions require photons with a narrow linewidth and are incompatible with short-pulse lasers \cite{Wang2022}.

\begin{figure*}[t]
	\centering
	\includegraphics[width=\linewidth]{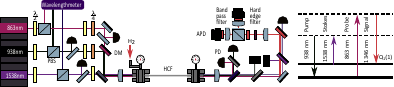}	
	\caption{Experimental setup. Two fields at \SI{938}{\nano\meter} (pump) and \SI{1538}{\nano\meter} (Stokes) are overlapped with the probe light at \SI{863}{\nano\meter} via dichroic mirrors (DM) and focused into the hydrogen-filled hollow core fiber (HCF). Behind the fiber, the converted light (signal, \SI{1346}{\nano\meter}) is separated from the input light. Two avalanche photon detectors (APDs) record the s- and p-polarized components after a polarizing beam splitter (PBS). The schematic diagram shows the underlying CSRS process.}
	\label{fig:setup}
\end{figure*}

The ARR-HCFs are a sub-group of the hollow-core photonic crystal fiber (HC-PCF), which are based on the guiding of the leaky mode through the eponymous hollow core \cite{Belardi2019}. Common implementations are named revolver-type fibers or single-ring PCFs and feature a small number of capillaries arranged around the hollow core at equal azimuthal distances within a jacket capillary \cite{Ando:19, Murphy:23}. The advantages of these fibers include an optical nonlinearity for the propagating mode combined with a tunable dispersion using a gas filling \cite{7225120}. Their properties are determined by the dimensions and arrangement of the holes. A combination of their outstanding polarization-maintaining property \cite{Taranta2020}, high damage threshold, low transmission loss, and tunable dispersion essential for phase matching in CSRS processes \cite{Russell2014} make these fibers ideal candidates for four-wave mixing processes \cite{Bahari:22,Afsharnia2024, Tyumenev2022-sp,Mridha:19,cwraman,PhysRevLett.99.143903}; see Ref.~\cite{novoa24} for a recent review.


\section{Experimental setup}

We perform frequency conversion from \SI{863}{\nano\meter} to \SI{1346}{\nano\meter}, where the choice of wavelengths is motivated by the potential conversion of the two entangled photons of the biexciton emission of indium arsenide (InAs)/gallium arsenide (GaAs) quantum dots \cite{Gao2012,Liu2019,Wang2019,Joecker2019} to the telecom O-band. The frequencies of two fields near \SI{938}{\nano\meter} and \SI{1538}{\nano\meter} are adjusted such that their beat note matches the vibrational transition at \SI{125}{\tera\hertz} within the $Q_1(1)$ branch of molecular hydrogen. The resulting nonlinear polarization is probed with the light field at \SI{863}{\nano\meter}, leading to emission at \SI{1346}{\nano\meter}. 

A schematic diagram of the experiment and the underlying conversion is shown in  Fig.~\ref{fig:setup}. The pump light at \SI{938}{\nano\meter} is provided by an amplified diode laser, while the \SI{1538}{\nano\meter} Stokes light originates from a diode laser combined with a fiber amplifier. These two lasers are stabilized at the MHz level using a wavelength meter, their power levels are adjusted to \SI{50}{\milli\watt}. In the following, pump and Stokes field are called pump fields. The probe photons are simulated through a low-power field derived from a diode laser operating at \SI{863}{\nano\meter}. The power levels of the three lasers are monitored, and their polarizations can be adjusted individually.

All beams are overlapped using dichroic mirrors and focused into the fiber with an achromatic lens. Each end of the fiber is glued into a centering piece, which in turn is mounted into a home-built, stainless steel, hermetically sealed, cylindrical high-pressure chambers using an O-ring seal. The vessels are tested up to \SI{700}{\bar}. AR-coated sapphire windows are sealed to the vessels using O-rings.
Behind the fiber, the converted photons at \SI{1346}{\nano\meter} are separated from the input light fields using dichroic mirrors and are guided to a polarizing beamsplitter (PBS) and two avalanche photodiodes (APD, ID Qube NIR, dead time \SI{1}{\micro\second}) with a quantum efficiency of \SI{12.5}{\percent}. Spectral filtering is achieved through bandpass filters (\SI{10}{\nano\meter} width and \SI{93}{\percent} transmission) and an edge filter at \SI{1300}{\nano\meter} with \SI{97.7}{\percent} transmission. 

\begin{figure}[tpb]
	\centering
	\includegraphics[width=\linewidth]{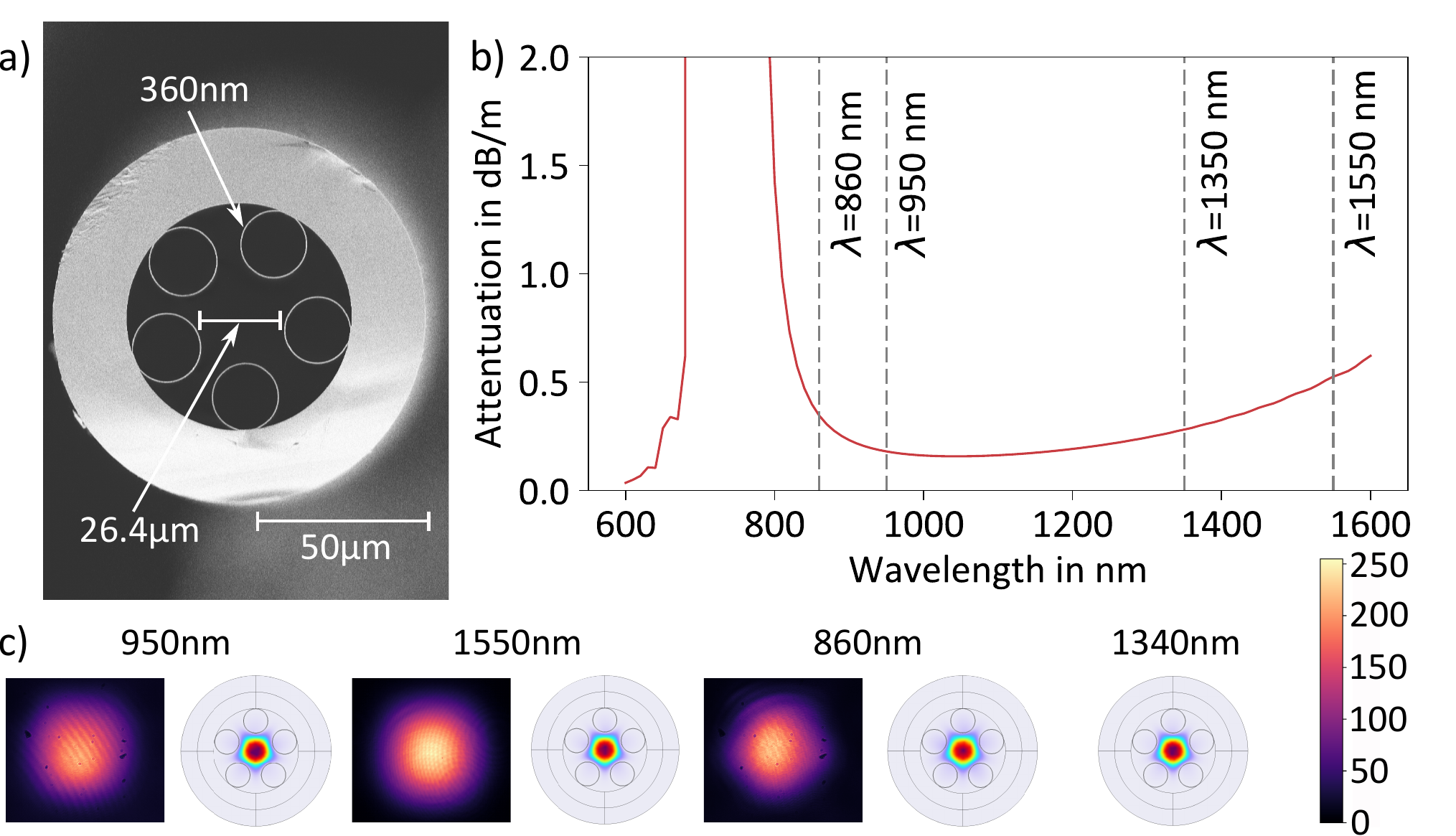}	
	\caption{a) SEM image of the fiber facet. b) Leakage loss as calculated  with \textit{Comsol}. c) Mode propagation inside the fiber calculated with \textit{Comsol}, showing the norm of the amplitude of the electric field inside the fiber, as well as images of the mode behind the fiber, for different wavelengths.}
	\label{fig:modes}
\end{figure}

We employ a single-ring ARR-HCF. The core has a diameter of \SI{26.4}{\micro\meter} and is surrounded by five capillaries with a wall thickness of \SI{360}{\nano\meter}, as shown in Fig.~\ref{fig:modes} a). The fiber is \SI{27}{\centi\meter} long. To calculate fiber properties such as the mode field diameter or the transmission region based on the chromatic dispersion, we use the model given by Zeisberger-Hartung-Schmidt \cite{fib6040068} to obtain the effective refractive index $n_\text{eff}$ inside the fiber core with small computational effort.

The coupling between core modes and capillary resonances \cite{Tyumenev2022-sp,fib6040068} leads to resonances, which in turn are associated with leakage loss and strongly depend on the wall thickness of the capillaries. The wall thickness is chosen such that the resonances do not coincide with the wavelengths used in the experiment.

Additionally, we use the finite element simulation package (\textit{Comsol}) for a simulation of the leakage loss and the Gaussian HE$_{11}$ fiber modes, shown in Fig.~\ref{fig:modes} b) and c).

In general, losses in hollow-core fibers can be attributed to various sources: leakage losses, imperfection losses, and material absorption of the fiber \cite{7225120}, as well as absorption by the hydrogen molecules \cite{Bahari:22}. Possible bend losses can also occur, but are negligible here. Leakage losses and material absorption \cite{7225120,Russell2014} are suppressed by the choice of the fiber with its resonance wavelengths, and imperfection losses only play a significant role for visible and ultraviolet wavelengths \cite{7225120}.

For the type of fiber used here, coupling efficiencies in excess of 90\% have been obtained \cite{zuba23}, but this value has been reduced to \SI{30}{\percent} in the high-pressure vessel after filling with hydrogen. We suspect that mechanical stress during pressurization non-reversibly damaged or tilted the fiber tip, which will be avoided in an upcoming design of the mounting structure.


\section{Frequency Conversion}\label{sec:freq}

\subsection{Role of the gas pressure} \label{sub:eff}
\begin{figure}[b]
	\centering
	\includegraphics[width=\linewidth]{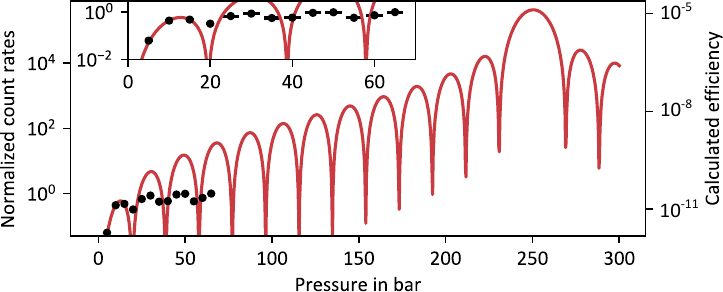}
         \caption{Pressure-dependent count rates (black circles) and calculated efficiency, see the text for details. The inset shows a zoom into the pressure region of up to 70 bar.}
	\label{fig:eff}
\end{figure}

The nonlinear conversion needs to fulfill the phase matching condition $\Delta \beta = -\beta_\text{938nm} +\beta_\text{1538nm}+\beta_\text{863nm}-\beta_\text{1346nm}$. We perform a calculation of the conversion efficiency in dependence of gas pressure. Starting from the propagation constant $\beta = 2 \pi \nu/c \cdot  n_{\text{eff}}(p,\lambda)$ and taking $n_{\text{eff}}$ in dependence of the pressure $p$ and the wavelength $\lambda$ from the Zeisberger model \cite{fib6040068}, we find an oscillatory behavior with a peak efficiency near \SI{250}{\bar}, shown as the solid curves in Fig.~\ref{fig:eff}.

In the experiment, we analyze the molecular resonance of the CSRS process  by scanning the relative detuning of the two pump fields. We find that the pressure-dependent lineshapes, linewidths, and collisional shifts \cite{Bischel86,Looi1978, Welsh1961} are identical to our previous work on a bulk hydrogen gas \cite{Aghababaei2023-pa,Hamer:24}. 

The conversion efficiency is pressure-dependent and scales as \cite{10.1063/1.5030335}
\begin{equation}\label{eq:I}
    I_\text{1346} \propto \lvert \chi^{(3)}(\omega) \rvert^2 L^2 \sinc^2\left(\frac{\Delta \beta (p) \cdot L }{2}\right) I_\text{938} I_\text{1538} I_\text{863},
\end{equation}
where $\chi^{(3)}(\omega)$ denotes the third-order nonlinear susceptibility, $L$ the interaction length, and $\Delta \beta$ is the previously mentioned pressure-dependent phase matching condition included in a squared $\sinc$ function. The normalized count rates as well as the calculated efficiency are shown in Fig.~\ref{fig:eff}.

In the regime of low pressures of up to about 20 bar, the data nicely follows the model. For the first maximum at \SI{12 \pm 1}{\bar} and the previously described experimental parameters, we obtain an internal conversion efficiency of \SI{1.8e-11}{} (relative efficiency \SI{7.2e-7}{\percent/\watt\squared}). In the current experiment, we refrained from a drastic increase in pump powers to not damage the fiber. 

After scaling by the power of the pump fields used, this is an improvement by two orders of magnitude compared to our previous work in a bulk gas \cite{Hamer:24}, achieved through the longer interaction length and the tighter confinement of the modes.

The count rate follows an oscillatory behavior with the same periodicity as predicted by the model. For higher pressures above about \SI{20}{\bar}, however, the measured efficiency falls significantly short of expectations and does not closely follow the model.
While the model does neglect subtle effects such as backcoherence \cite{Tyumenev2022-sp}, higher order Stokes shifts \cite{Mridha:19}, and intermodal conversion \cite{Mridha:19}, we here discuss two explanations for the saturation in conversion efficiency at large gas pressure.

Firstly, we experimentally observe significant gas losses occurring at pressures above \SI{20}{\bar}, leading to a pressure gradient along the fiber. We therefore only use data up to \SI{20}{\bar} for the fitting routine. The inhomogeneous pressure, possibly along with the gas flow through the fiber, disrupts coherence along the length of the fiber and thus limits the efficiency. This observation is verified in simulations. In future work, uniform pressure within the fiber will be established by placing the entire fiber inside a high-pressure vessel.

Secondly, and conceptually more relevant, the intensity provided by the two cw pump fields might be insufficient to generate a sufficiently strong coherent wave of optical phonons which behave as stationary oscillators \cite{Ziemienczuk:12} and drive the conversion process. Previous experiments \cite{Tyumenev2022-sp,novoa24} employed pulsed lasers with substantially higher peak intensities. Additionally, in the approach described here, the Stokes photon was not generated by stimulated Raman scattering (SRS), but was instead provided by a laser as well. A redesign of the fiber mount will allow us to improve on the fiber coupling efficiency, which in turn will allow us to increase the pump fields and elucidate the build-up of coherence with increasing pump power.


\subsection{Preservation of polarization}
Quantum frequency conversion requires preservation of the incoming photon state, particularly the polarization. Since the hydrogen gas itself is isotropic, the reference frame of the system is determined by the light field. We probed the preservation of polarization by measuring the projection of the output state for both linear and circular polarizations. To this end, the polarization of the probe light is varied, and a polarizing beamsplitter (PBS) separates the V and H polarizations of the converted light onto two separate detectors. Silver mirrors are used to maintain the polarization, and the optics were characterized by means of a test beam near \SI{1310}{\nano\meter}.

\begin{figure}[tpb]
	\centering
	\includegraphics[width=\linewidth]{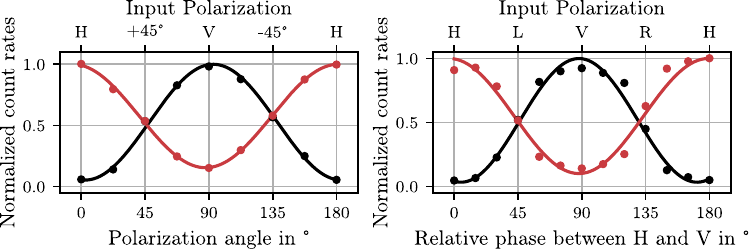}
         \caption{Preservation of polarization, demonstrated here for linear (left) and circular (right) polarization. Black data shows vertical polarization, red data shows horizontal polarization at the detector.} 
	\label{fig:HC_HD}
\end{figure}

For all values of linear and circular polarization, the converted light faithfully follows the input polarization; see Fig.~\ref{fig:HC_HD}. To calculate the fidelity, sine functions are fitted to the background-corrected data. We obtain a fidelity of \SI{96}{\percent} for linear polarization and \SI{84.5}{\percent} for circular polarization, leading to an overall fidelity above \SI{0.91}{}. This value can be fully explained by imperfect polarization optics and allows us to conclude that the fiber does not deteriorate the polarization \cite{Taranta2020}.


\subsection{Background noise}

Frequency conversion in crystals is often plagued by undesired Raman scattering into the spectral band of the converted photons. These processes are driven through the crystal material, and there is justified hope that such processes are suppressed or removed entirely in the case of a gas-filled hollow-core fiber where the glass structure has minimal overlap with the light fields. Indeed, we do not observe any background near the target wavelength of 1346 nm that would originate from either of the two pump fields or their combination.

When increasing the power of the 863-nm probe field to the level of a few 10 mW, we do observe weak emission near 1346 nm that scales linearly with both the power of the 863-nm field and the pressure of the hydrogen gas. This emission is attributed to spontaneous Raman scattering in hydrogen \cite{BOYD20081}. At a pressure of 5 bar, we obtain an internal spontaneous conversion efficiency of \SI{1.3e-12}{}, corresponding to an efficiency of \SI{7e-15}{}/(cm bar). The coherent four-wave mixing process can be ten orders of magnitude stronger than this spontaneous process \cite{Tyumenev2022-sp}.


\section{Conclusion} 
To conclude, we have demonstrated frequency conversion in high-pressure hydrogen entrapped inside a hollow-core fiber, which increases mode overlap and interaction length by orders of magnitude. Different from all related studies thus far \cite{Tyumenev2022-sp}, we employ continuous-wave pump fields which do not reduce the bandwidth of the signal photons, but allow for broadband conversion of photons with large coherence time. In future work, we will increase the conversion efficiency through a re-design of the high-pressure vessel and fiber mounts to facilitate higher incoupling efficiency, better power handling, and increased pressure stability.

\section{backmatter}
\subsection{Funding}
We acknowledge funding by Deutsche Forschungsgemeinschaft DFG through grant INST 217/978-1 FUGG and through the Cluster of Excellence ML4Q (EXC 2004/1 – 390534769), as well as funding by BMBF through the QuantERA project QuantumGuide.

\subsection{Acknowledgments}
We thank all members of the Cluster of Excellence ML4Q and the QuantumGuide collaboration, as well as Philipp Hänisch, for stimulating discussions. 

\subsection{Disclosures}
The authors declare no conflicts of interest.

\subsection{Data Availability Statement}
Data underlying the results presented in this paper are not publicly available at this time but may be obtained from the authors upon reasonable request.


\bibliography{bib}

\end{document}